\documentclass{article}
\usepackage[dvips]{graphicx}
\usepackage{epsfig}
\usepackage{longtable}
\usepackage{float}
\begin{document}

\title{   \centerline{Localization of water monomers inside ice-like clusters. }  }
\author{ V.L.Golo$^{*}$ and S.M.Pershin${**}$  \vspace{3mm }\\
         $^{*}$ The Lomonosov Moscow State University, \\
	 Department of Mechanics and Mathematics,\\
	 e-mail: voislav.golo@gmail.com;   \vspace{3mm } \\
	 $^{**}$ The Prokhorov General Physics Institute,\\
	 Russian Academy of Sciences, \\
	 e-mail: pershin@kapella.gpi.ru.
}

\maketitle


\begin{abstract}
On the basis of the experimental data we suggest that water monomers could be trapped in channels running through
ice-like clusters  in water. Our argument relies on  a simple model that describes
 the motion of a  dipole particle  inside a channel in the presence of an electric field
with linear gradient. The model admits of both finite and infinite regimes of
motion so that the finite one could correspond to the particle being confined to a channel.
\end{abstract}

\section{ Introduction: Water monomers trapped in channels of ice-like structures.}
It was W.C. R\"ontgen, \cite{roentgen} who suggested the presence of ice-like structural fragments in liquid water, and  put forward
the hypothesis of two fluid structure of water in equilibrium. Later, O.Ya.~Samoilov, \cite{samoilov}, and L.Pauling, \cite{L. Pauling},see also \cite{frenkel}, extended the R\"ontgen model by introducing the concept of clathrate, low density and high density water. The idea has obtained  experimental support that provides evidence for the existence of two kinds of water. Thus, using the technique of four-photon spectroscopy  we have observed,
for the first time to the best of our knowledge,
the rotational resonances in water due to monomers $H_2O, \; H_2O_2$ and $OH$ \cite{p1}, \cite{p2}.  It is important that
the  mobility of monomers could  explain  the high  permeation of water channels, in biological membranes, \cite{murata}.
As to the ice-like structural fragments,  recently, Nielsson et al,\cite{niels}, using  x-ray spectroscopy,  have found  experimental evidence
in favor of their existence. They  visualized them  as  cluster structures, of  several tens of \AA{}   and hexagonal ice symmetry $1h$.

Considering water  as a mixture of ice-like clusters and monomers, makes for better understanding the physics of water.  Nevertheless,
there are still questions  about  the nature of coexistence of the monomers and the ice-like clusters.
Specifically, the penetration  of the $H_2 O$ monomers  through  channels of the ice-like clusters needs  exploring.
The key to the problem is provided by the study of beams of fast particles in crystalline media, effected many years ago for the needs of nuclear physics, \cite{tulinov}. The main point is that the velocity  of a particle is greatly enhanced if the beam  is directed along a crystalline axis  providing a kind of channel. Thus, one can claim the channeling effect, which appears to be of a significance reaching outside  nuclear physics. In fact the problem is the old one.
Years ago,  Ya.I. Frenkel, \cite{frenkel}, considered the motion of molecules and ions in channels inside crystalline structures,
see also \cite{zacepina}.

The observation of rotational modes of water monomers suggests the existence of cavities that could accommodate
the motion on a time scale larger than that due to the switching of hydrogen bonds in water, that is $1 - 2 \; ps$ according to paper \cite{Nibbering}.  The channels inside a cluster with hexagonal ice 1h structure,
of diameter $ 5.7$ \AA, could serve such cavities. It is worthwhile to notice that
similar  equipotential cavities can be provided through the freezing of water in a cryogenic matrix
formed by argon, \cite{xavier}, and carbon nanotubes of diameter $14$ \AA, \cite{kolesnikov}.
It is reported,\cite{kolesnikov},  that  molecules of water preserve their mobility  in carbon nanotubes down to temperatures  $\approx~8 K$.

The main point about the dynamics of  water monomers is that  they could remain trapped in crystalline channels
instead of forming complexes of hydrogen bonds  and quitting the channels.
In this paper we are considering a simple qualitative model that could accommodate the phenomenon.
Our arguments essentially rely on the interplay of translational motion of a monomer
and its rotational dynamics caused by the dipole moment of  water molecule .
\begin{figure}[H]
    \center{    \includegraphics[height=7cm,width=17cm]{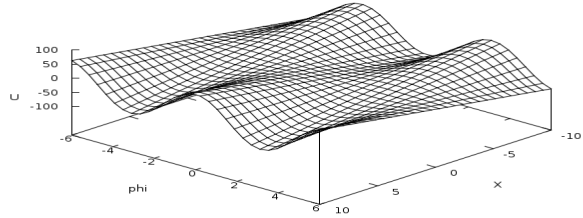}
           }
       \caption{Potential  $U = p \, E(x) \, cos(\varphi)$}
       \label{fig:Potential}
\end{figure}

\section{A quasi-classical model for the motion of monomer }
We shall draw a qualitative picture of  the  motion of a monomer as the dipole in the following potential 
$ U = p \, E(x) \, cos(\varphi)$.  Potential surface is presented in Fig.\ref{fig:Potential}  .
To that end we shall consider the latter  as a two-dimensional rotator
moving in direction of axis-$x$ in an external electric field, $E(x)$, which mimics constraints imposed  by the channel.
Thus, its configuration  is specified by angle $\varphi$ describing its rotation and coordinate $x$ its position in axis-$x$.
Its dynamics is described by angular momentum $L$ and momentum $P$.
It is important to choose the right characteristic scales of the system. We shall take:\\
(1) time scale $\tau = 10^{-14} \; sec$   corresponding to  rotations  of water molecules; \\
(2) spatial scale $r = 3 \cdot 10^{-8} \; cm$, close to the size of a water molecule; \\
(3) mass scale $3.2 \cdot 10^{-23}$ corresponding to a water molecule;  \\
(4) dipole moment $1 D = 10^{-18 } \; CGS$; \\
(5) electric field $10 \; kV /cm$.\\
Hence  we infer: \\
(6) velocity and momentum scales  $3 \cdot 10^6 \; cm / sec$, $10^{-16} \; gr \, cm \, sec^{-1}$, respectfully; \\
(7) moment of inertia $\approx 3 \cdot 10^{-38} gr \, cm^2$.\\
We shall assume that the external electric field, $E(x)$, is linear in $x$
$$
  E(x) = A \, x \; + \; W
$$
The scales are conducive to the use of  quasi-classical approximation and  numerical simulation.
In fact, for the above characteristic scales we have the de Broglie wave length
$$
    \lambda = \frac{\hbar}{P} \approx 10^{-11} \;  cm .
$$
Assuming the size of a channel $ {\cal L}\approx 10$ \AA, we get the parameter of quasi-classical expansion
$$
       \mu =  \frac{\lambda}{{\cal L}} \approx 10^{-4},
$$
that is sufficiently small. The usual constraint imposed on the interaction potential, \cite{bloch},
$$
     \left | \frac{\partial U(\overline{x})}{\partial \overline{x}}  \right |  \gg
     \frac{1}{2} \left |  \frac{\partial^3 U(\overline{x})}{\partial \overline{x}^3}  \right | \; \overline{\Delta x^2},
$$
which means that the potential is smooth enough at the  de Broglie  length, is valid for the interaction between the dipole moment and the external field
$$
    U(x, \varphi) = p \cdot \ E(x) \; cos(\varphi)
$$
for $E(x)$ is linear in $x$. The similar requirement for the rotational motion is also verified.  We have the dimensional expansion parameter
$$
    \mu_{\varphi} = \frac{\hbar}{L} \approx3 \cdot 10^{-4}, \qquad L = I \dot {\varphi} \approx 3 \cdot 10^{-24} \; erg \cdot sec
$$

It should be noted that the requirements indicated above are not satisfied at return points where the size of de Broglie length rapidly changes.
For example, this is the case even of the harmonic oscillator. Thus, the conventional quasi-classical approximation breaks down, \cite{bloch}.  At these points one can change the x-representation for the p-representation, which, from the formal mathematical point of view, amounts to  the Fourier transform.
But the advanced theory, \cite{maslov}, \cite{sternin}, of the quasi-classical approximation claims that one can still use the classical equations of motion in conjunction with the Bohr-Sommerfeld quantization condition
$$
  \int \; p \, dq = 2 \, \pi \, n \, \hbar,  \qquad \mbox{where } n \mbox{  is an integer},
$$
provided the gradient of the Hamiltonian is not degenerate
\begin{equation}
      \left( \frac{\partial H}{\partial x} \right)^2 \; + \; \left( \frac{\partial H}{\partial p} \right)^2 \ne 0,
      \label{eq:degen}
\end{equation}
where $x, \; p$  are coordinates and momentum of the system. It should be noted that we  are not constructing the wave function at turning points,
but only find the quasi-classical trajectory by means of the classical Hamiltonian equations. It is the situation of the 'old' quantum theory by Niels Bohr.

The arguments given above enable us to describe the dynamics of rotator within the framework of Lagrangian mechanics.
The Lagrangian function reads
\begin{equation}
\label{eq:lagrange}
   {\cal L} = \frac{I}{2} \dot{\varphi}^2 \; + \;  \frac{m}{2} v^2 \; - \; p \cdot \ E(x) \; cos(\varphi)
\end{equation}
Here $I = 3 \cdot 10^{-38} \; gr \, cm^2$ is the moment of inertia of rotator equal by orders of magnitude to that of a molecule of water;
$m = 3.2 \cdot 10^{-23} \; gr$ is the mass  of rotator, close to the mass of a molecule of water; $p$ is the dipole moment of a water molecule.
We take the electric field of the form
\begin{equation}
\label{eq:elfield}
      E = A \, x \; + \; W
\end{equation}
It is worth noting that in our case both terms in the kinetic energy are  $\approx 10^{-10} \; erg$, whereas the potential energy is smaller by two orders of magnitude,
that is $\approx 10^{-12} \; erg$. Considering the shape of the potential energy, it is easy to come to the conclusion
that  Lagrangian function (\ref{eq:lagrange}) admits both finite and infinite regimes of motion.  The equations of motion corresponding to Lagrangian function (\ref{eq:lagrange}) read
\begin{eqnarray}
\label{eq:rotmotion}
      m \, \ddot x & = &  p \; \frac{d  E}{d x} \; cos(\varphi ) \\
      I \, \ddot \varphi & =   - &  p \; E(x) \; sin(\varphi) \nonumber
\end{eqnarray}
\begin{figure}[H]
    \center{\includegraphics[height=13cm,width=15cm]{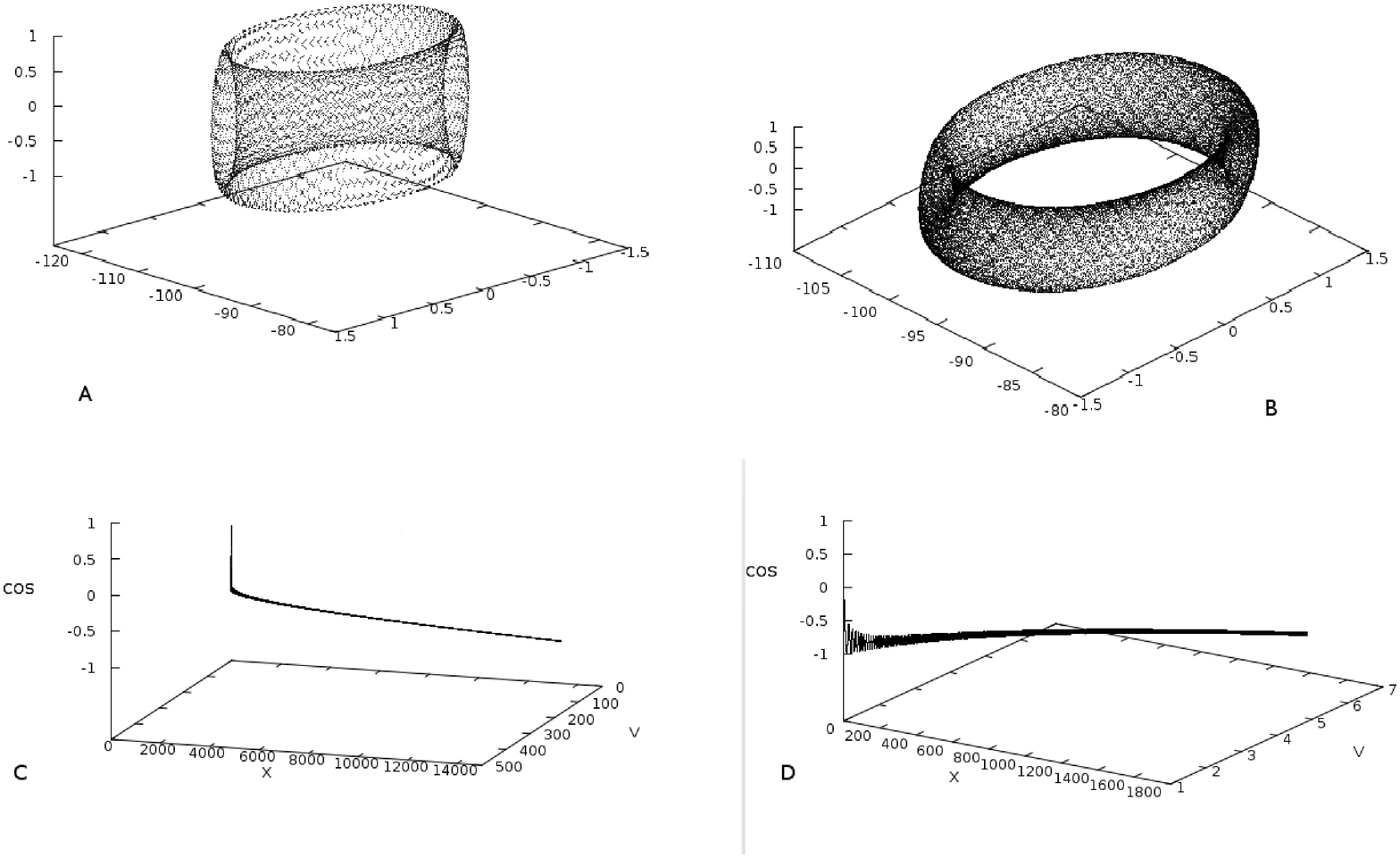}}
        \caption{ Motion of a dipole particle represented in the window given by the variables:
	          $X \quad$  x-coordinate of the rotator;
	          $V \quad$   velocity of the rotator  along x-axis;
		  $COS \quad$  $cos(\varphi)$ where $\varphi$ is the phase of rotator.
                  Dipole moment and rotator mass  $p = 1.8 D$, and $32 \cdot 10^{-24} \; gr$.
		  Moment of inertia $I = 3 \cdot 10^{-38} \; gr \, cm^2$.
		  Inserts {\bf A, B} and {\bf C, D} correspond to the confined and the infinite motions, respectfully.
                }
         \label{fig:localization}
\end{figure}
Equations (\ref{eq:rotmotion}) have only one integral of motion, the energy, and therefore one cannot solve it in a finite form,
that is by writing down its solution by means of integrals. We have to employ numerical simulation for studying it.
The key problem is the wise choice of visual representation of numerical results. In this respect, it should be noted that equations (\ref{eq:rotmotion}) are two equations of second order, and therefore their phase space is four dimensional comprising coordinates $x, \; \varphi$ and velocities $\dot x, \; \dot \varphi$.  We have used variables $x, \; v = \dot x$ and $cos( \varphi)$ that provide a kind of three-dimensional window on the four-dimensional phase space of equations (\ref{eq:rotmotion}). It is important that due to the energy integral the values of angular velocity, $\dot \varphi$, are bounded, and therefore
we can infer the character of motion from the picture of a trajectory in the above window. The results of the simulation are illustrated in Fig.\ref{fig:localization}. We have employed the following values for  coordinates and fields:
\begin{itemize}
 \item[\bf A]        initial velocity, phase, and angular velocity{\bf:}
		  $3 \cdot 10^6 \; cm / sec, \quad \varphi = 0, \quad 4 \cdot 10^{14} \; Hz$,  respectfully.
		  Background field and field gradient
		  $W = 5 \; kV / cm$ and  $A = 500 \; kV / cm^2$.
		  Period of time considered $2 \cdot 10^{-11} \; sec$.
		  Amplitude of the particle's oscillation in $x$ several tens of \AA.
  \item[\bf B]       initial velocity, phase, and angular velocity{\bf:}
		  $3 \cdot 10^6 \; cm / sec, \quad \varphi = 0, \quad 4 \cdot 10^{14} \; Hz$,  respectfully.
		  Background field and field gradient
		  $W = 5.3 \; kV / cm$ and  $A = 500 \;  kV / cm^2$.
		  Period of time considered $ 10^{-10} \; sec$.
		  Amplitude of the particle's oscillation in $x$ several tens of \AA.
  \item[\bf C]       initial velocity, phase, and angular velocity{\bf:}
		  $ 3 \cdot 10^6 \; cm / sec, \quad \varphi = 1.7 \; rad, \quad  10^{13} \; Hz$,
		  Background field and field gradient
		  $W = 10 \; kV / cm$ and  $A = 10 \;  kV / cm^2$.
		  Infinite motion.
  \item[\bf D]        initial velocity, phase, and angular velocity{\bf:}
		  $ 3 \cdot 10^6 \; cm / sec, \quad \varphi = 1.7 \; rad, \quad  10^{12} \; Hz$,
                  Background field and field gradient
		  $W = 0.1\; kV / cm$ and  $A = 0.1 \;  kV / cm^2$.
		  Infinite motion.
\end{itemize}

The motion being finite, that is  the rotator confined to a finite region of x-axis,
depends on values of initial  coordinates and velocities, and  values of the background field, $W$, and the field gradient $A$.
The change of dynamical regimes has the threshold nature, so that small changes of fields and initial position may,
generally, result in  different regimes of motion.

\section{Conclusions}
 The numerical analysis of the dynamics of a rotator indicates that there exist regimes corresponding to the confinement of rotator
 to  finite regions of phase space. The essential point is that the motion takes place in {\it a field that increases linearly
 as regards the spatial coordinate},  when a charged particle would move neglecting any boundaries. Thus, the finite motion
 is due to the interplay of translational and rotational degrees of freedom. In this respect it strongly resembles Maxwell's pendulum.
 The latter consists of a flywheel and two wires wound round the flywheel axis in the same direction and connected with a horizontal support.
 When released the flywheel goes down rotating round its axis due to the attached wires,
 until it arrives at the lowest point allowed by unwinding wires. Then it goes upwards rewinding in the opposite direction.
 The equations of motion   are similar to those of the one dimensional rotator,
 so that one can consider  Maxwell's pendulum as a kind of mechanical model for the confinement of a particle in channel.

 The model studied in this paper is the  qualitative one.  It provides the  argument in favor of the claim
 that a monomer of water may be confined to a channel, or cavity,  inside an ice-like cluster of liquid water  hydration layers in the vicinity of "membrane water channels", \cite{murata}.
 Specifically, we suggest that the $H$ and $H_30^{+}$ ions, which always exist in water, \cite{samoilov}, \cite{zacepina},
 could produce a gradient electric field  that may result in the channeling effect.

The useful discussion with B.Yu.Sternin is gratefully acknowledged.

\end{document}